\documentstyle[aps,multicol,eqsecnum,epsf]{revtex}

\begin{document}
\draft

\title{Continuous Avalanche Segregation of Granular Mixtures in Thin
Rotating Drums} \author{Hern\'an A. Makse} \address{ Schlumberger-Doll
Research, Old Quarry Road, Ridgefield, CT 06877}
\date{Phys. Rev. Lett., {\bf 83}, 3186 (1999)}
\maketitle
\begin{abstract}
We study segregation of granular mixtures in the continuous avalanche
regime (for frequencies above $\approx 1$ rpm) in thin rotating drums
using a continuum theory for surface flows of grains.  The theory
predicts profiles in agreement with experiments only when we consider
a flux dependent velocity of flowing grains.
We find the segregation
of species of different size and surface properties, with the smallest
and roughest grains being found preferentially at the center of the
drum.
For a wide difference between the species we find a complete
segregation in agreement with experiments.
In addition, we predict a transition to a smooth segregation regime---
with an power-law decay of the concentrations as a function of radial
coordinate--- as the size ratio between the grains is decreased
towards one.

\end{abstract}

\begin{multicols}{2}

Mixtures of grains differing in size, shape or density segregate in
vibrated containers \cite{review}, in two and three-dimensional
rotating drums \cite{nakagawa,zik,hill,khakhar,duran,cantalaube}, and
by pouring the mixture into a heap \cite{makse1,hans}.  For instance,
axial band segregation appears in long rotating cylinders filled with
mixtures of grains \cite{nakagawa,zik,hill}, while radial segregation of
granular mixtures is observed in thin ($\approx 10$ grain size)
rotating drums:
%
large grains are usually being found near the outer region of the
drum, and small grains near the center of the drum
\cite{duran,cantalaube}.

Theoretically, the segregation of mixtures in rotating drums has been
studied by Zik {\it et al.}  \cite{zik} using a continuum approach
based on mass conservation to treat band formation in 3-D
drums. Recently, Khakhar {\it et al.}  \cite{khakhar} predicted the
concentration profiles in 2-D rotating drums filed with mixtures
composed of equal-size grains of different density. However, there is
no general analytical solution for segregation of grains in 2-D
rotating drums in the case of mixtures of arbitrary size, and shape or
friction properties.  In this Letter, we study analytically this case
focusing on the continuous avalanching regime, where a steady state is
observed.  Based on a continuum theory of surface flows of grains,
we obtain predictions for the profile of the rolling species which
agree quantitatively with experiments \cite{nakagawa,khakhar} when we
take into account the velocity profile of the flowing grains.

We find that the small and rough grains occupy the center of the drum
while the large and rounded grains segregate to the bottom of the
drum.  According to the size ratio and the different angles of repose
of the species, there are two distinct regimes.  For large size ratios
our solution shows almost complete segregation of the mixture with a
small region of mixing near half of the drum radius.  The surface
concentrations of grains behave exponentially on radial coordinate and
segregation is characterized by a typical length scale inversely
proportional to the size ratio between the grains.  These findings
agree with the experiments of Cl\'ement {\it et al.}  \cite{duran}.
On the other side, when the species differ slightly, we
predict a slow power-law decay of the concentrations of grains at the
surface, with no characteristic length scale associated to
segregation.


In the continuous avalanching regime,
grains on the free surface flow down steadily due to gravity forming a
film of rolling species, while the bulk of the material rotates as a
solid body. Thus, it is plausible to describe the system only by its
surface properties \cite{bcre,pgg,bdg,makse2}.

\begin{figure}
\centerline{
\vbox{ \hbox{\epsfxsize=6.cm
\epsfbox{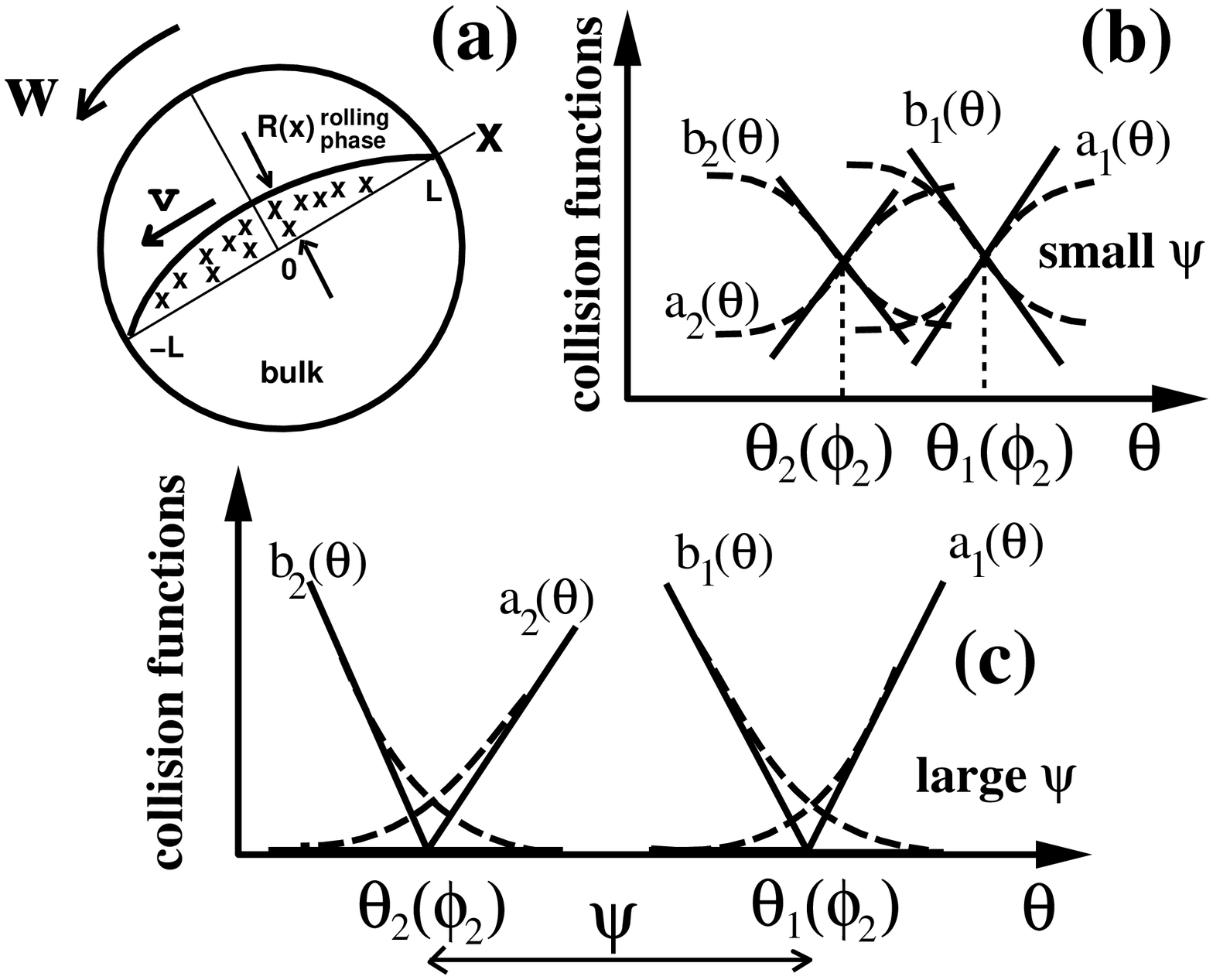}}
}}
\narrowtext
\caption{}
\label{drum}
\end{figure}

We consider two local coarse-grained equivalent thicknesses of the
species in the rolling phase $R_{i}(x, t)$, with $i =1,2$ [i.e. the
total thickness of the rolling phase multiplied by the local volume
fraction of the $i$ grains in the rolling phase at position $x$ on the
surface of the bulk, Fig. \ref{drum}(a)].  The total thickness of the
rolling phase is then $R(x,t) \equiv R_1(x,t) + R_2(x,t)$.  The static
phase rotating as a bulk is described by the local angle of the
surface height, $\theta(x,t)\equiv-\partial h(x,t)/ \partial x$
[$h(x,t)$ is the height], and the surface fraction of static grains,
$\phi_i(x,t)$, of type $i$ at the surface of the bulk.  The equations
of motion for the rolling species are \cite{bdg}

\begin{mathletters}
\label{bdg-eq}
\begin{equation}
\label{bdg-r}
\frac{\partial R_i(x,t)}{\partial t}= v(x,t) \frac{\partial
R_i}{\partial x} +
\Gamma_i,
\end{equation}
and the surface concentrations  of static grains 
are given by
\begin{equation}
\phi_i(x,t) (\frac{\partial h}{\partial t} - w x) = - \Gamma_i,
\label{bdg-phi}
\end{equation}
\end{mathletters}
\noindent
and
$ \phi_1 + \phi_2 =1$.
Here $w$ is the rotation frequency of the drum. The downhill
convection velocity of the rolling species $v(x,t)$ is usually taken
as a constant \cite{zik,bcre,pgg,bdg,makse2}. However, experiments on granular
flows in  drums,  inclined chutes and sandpiles 
\cite{nakagawa,savage,makse3} indicates
that the velocity  varies linearly with the vertical 
position, $y$, 
 of the particle in the
flowing layer, i.e. $v(y)=2\alpha y$, where $\alpha$ is the shear rate.
In our continuum approach we take a height-averaged
velocity dependence 
of the form \cite{makse3}: $v(x,t)=1/R(x,t) \int_0^R v(x,y,t) dy=
\alpha ~R(x,t)$.
We will show that the flux dependence of the velocity plays a crucial
role in explaining the experimental results.

The interaction term $\Gamma_i$ takes into account the conversion of
static grains into rolling grains and vice versa, and is \cite{bdg},
[$i=1,2$, ($j\ne i$)]
\begin{equation}
\Gamma_i 
\equiv a_i(\theta) \phi_i R_i - b_i(\theta) R_i + x_i(\theta)  \phi_i R_j.
\label{gamma}
\end{equation}

The interaction term is a function of the collision functions
contributing to the rate processes. Amplification, $a_i(\theta)$: 
an $i$ static grain is converted into a rolling grain due to a
collision by a rolling grain $i$.  Capture, $b_i(\theta)$: an $i$
rolling grain is converted into a static grain. Cross-amplification,
$x_i(\theta)$: the amplification of a $j$ static grain due to a
collision by an $i$ rolling grain.



We consider the geometry of a two-dimensional rotating drum of radius
$L$ [$-L\le x \le L$, Fig. \ref{drum} (a)].  We study the continuous
avalanche regime where steady state profiles are observed in the drum,
and we seek a solution $\partial h(x) /\partial t = 0$, $\partial
R_i(x)/\partial t = 0$, with boundary conditions $R_i (x=-L) = R_i
(x=L) =0.$

\begin{figure}
\centerline{
\vbox{ \hbox{\epsfxsize=6.cm
\epsfbox{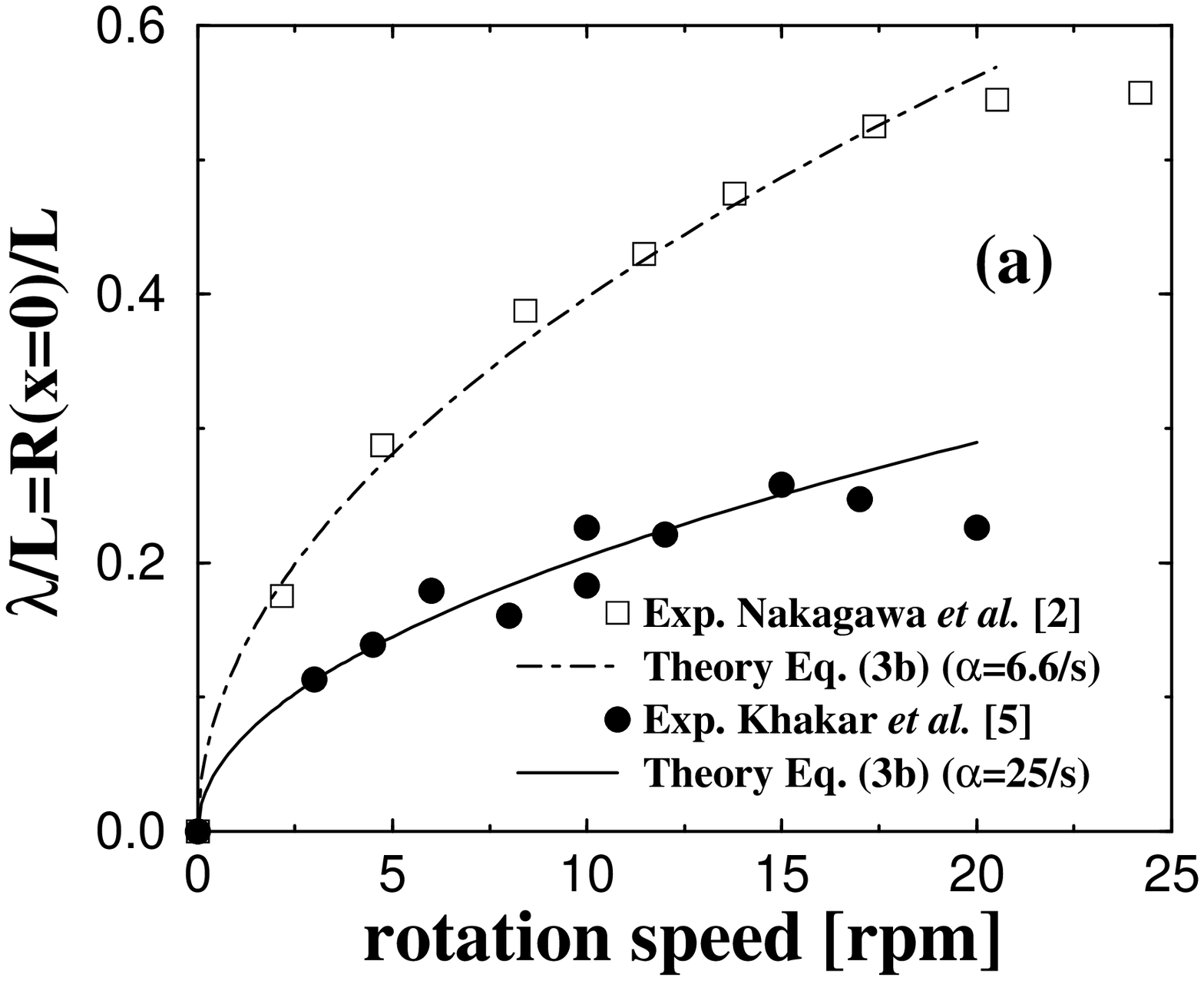}}
\epsfxsize=6.cm
\epsfbox{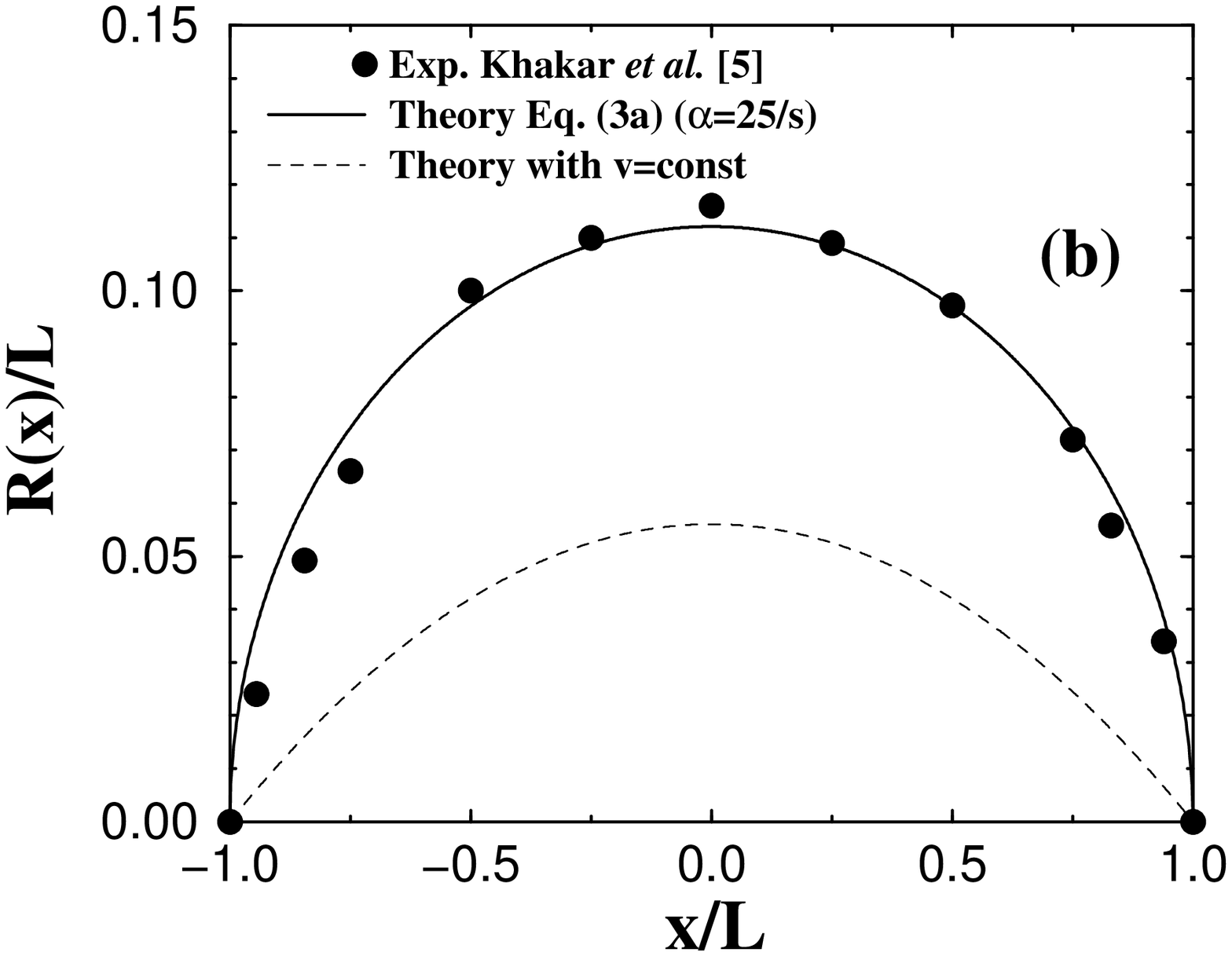}
}}
\narrowtext
\caption{}
%
\label{exp}
\end{figure}

The profile of the total thickness of the rolling phase $R(x)$ and the
thickness of the rolling layer at the center of the drum $\lambda
\equiv R(x=0)$ are obtained from (\ref{bdg-eq}):
\begin{mathletters}
\begin{equation}
R(x)=  \sqrt{\frac{w }{\alpha}} ( L^2 - x^2)^{1/2},
\label{total}
\end{equation}
\begin{equation}
\lambda\equiv R(x=0)=\sqrt{\frac{w}{\alpha}}L
\label{lambda}
\end{equation}
\label{total-lambda}
\end{mathletters}
The square root dependence in (\ref{total}) differs from the parabolic
profile found in \cite{pgg,zik}.
This latter result \cite{pgg,zik} is obtained by assuming
$v=$const. Solving Eqs. (\ref{bdg-eq}) for this case gives: $R(x)=
w/(2v) (L^2-x^2)$, and $\lambda=w L^2/(2v)$.  In
Fig. \ref{exp}(a) we plot $\lambda(w)$ and compare with the digital
photography experimental data of Khakhar {\it et al.} \cite{khakhar}
for spherical sugar balls of 1.8 mm size, and the MRI experiments of
Nakagawa {\it et al.} \cite{nakagawa}.  We clearly see that
$\lambda(w)$ is consistent with a square root dependence on $w$ as in
(\ref{lambda}) and not with a linear dependence on $w$ obtained by
assuming $v=$const \cite{deviations}.
In Fig. \ref{exp}(b) we plot the theoretical prediction (\ref{total})
for $R(x)$, as well as the prediction obtained when $v=$const, and
compare with the experimental results of \cite{khakhar}.  Since we use
$\alpha=25/$s obtained from Fig. \ref{exp}(a), and $w=2\pi f$ with
$f=3$ rpm from the experiments, the comparisons in Fig. \ref{exp}(b)
are made without tuning of parameters.  We again see an excellent
agreement between experiment and theory only when the flux-dependent
velocity is taken into account,
while the prediction
based on the assumption of a constant velocity profile 
gives lower values of $R(x)$ \cite{parabolic}.
This result is in agreement with the numerical analysis
of \cite{khakhar} for simple shear flows.

%

To describe the system composed of a binary mixture of grains, we
propose below a set of collision functions valid in the limiting cases
of a wide and small difference between the species.  We first treat
the case where the difference in size and angles of repose is not too
large (we will quantify this later).  In this case the region of
interest is concentrated in a small region around the angles of
repose, a fact that allows us to linearize the collision functions in
the vicinity of the angles of repose. We propose [Fig. \ref{drum} (b)]

\begin{equation}
\begin{array}{rll}
a_i(\theta)\equiv x_i(\theta) &\equiv& C + \gamma ~ [\theta(x)-
\theta_i(\phi_j)], \\ b_i(\theta)&\equiv& C - \gamma ~ [\theta(x)
-\theta_i(\phi_j)].
\end{array}
\label{linear}
\end{equation}
Here, $\gamma \approx 23/$s \cite{makse3} and $C$ are collision
rates.  In the continuous avalanche regime we have $w>\gamma
\delta^2/2$, where $\delta$ is the difference between the maximum
angle of stability and the pure angle of repose of the species
\cite{pgg}.  We note that the collision functions (\ref{linear})
differ from the ones proposed in \cite{bdg} in the crucial fact that
the generalized angle of repose, $\theta_i(\phi_j)$, of a rolling
grain type $i$ depends on the composition of the surface $\phi_j$
\cite{makse2}:
\begin{equation}
\label{dependence}
\begin{array}{rcl}
\theta_1(\phi_2) &=& m \phi_2 + \theta_{11},\\
\theta_2(\phi_2) &=& m \phi_2 + \theta_{21} = - m \phi_1 + \theta_{22}.
\end{array}
\end{equation}
The limiting cases are $\theta_{ij}$ = $\theta_i(\phi_j=1)$,
$m\equiv \theta_{12} - \theta_{11} = \theta_{22} - \theta_{21}$, and
the difference $\psi = \theta_1(\phi_2) - \theta_2(\phi_2) $ is
independent of the concentration $\phi_2$, then $\psi = \theta_{11} -
\theta_{21} = \theta_{12} - \theta_{22}$.

The angle $\psi$ is proportional to the degree of
difference in size and shape between the species, and therefore
determines the degree of segregation; the larger $\psi$, the stronger
the segregation.  If the species $1$ are the smallest then
$\theta_1(\phi_2) > \theta_2(\phi_2)$ for any composition of the
surface $\phi_2$, and also we have $\theta_{12} > \theta_{21}$.  For
mixtures of grains with different shapes or friction coefficients we
have $\theta_{11}\neq\theta_{22}$, and $\theta_{12} = \theta_{21}$ if
the species have the same size.  If $\theta_{11}$ is the repose angle
of the pure round species and $\theta_{22}$ is the repose angle of the
pure rough species, then $\theta_{11}<\theta_{22}$.  If the species
have the same shape or friction coefficients then
$\theta_{11}=\theta_{22}$.  When the size ratio is close to one
($d_2/d_1 < 1.5$ \cite{makse1,hans}, $d_i$ size of grain $i$), the
angle $\psi$ is small and then we linearize the collision functions as
in (\ref{linear}).

\begin{figure}
\centerline{
\vbox{ \hbox{\epsfxsize=6.cm
\epsfbox{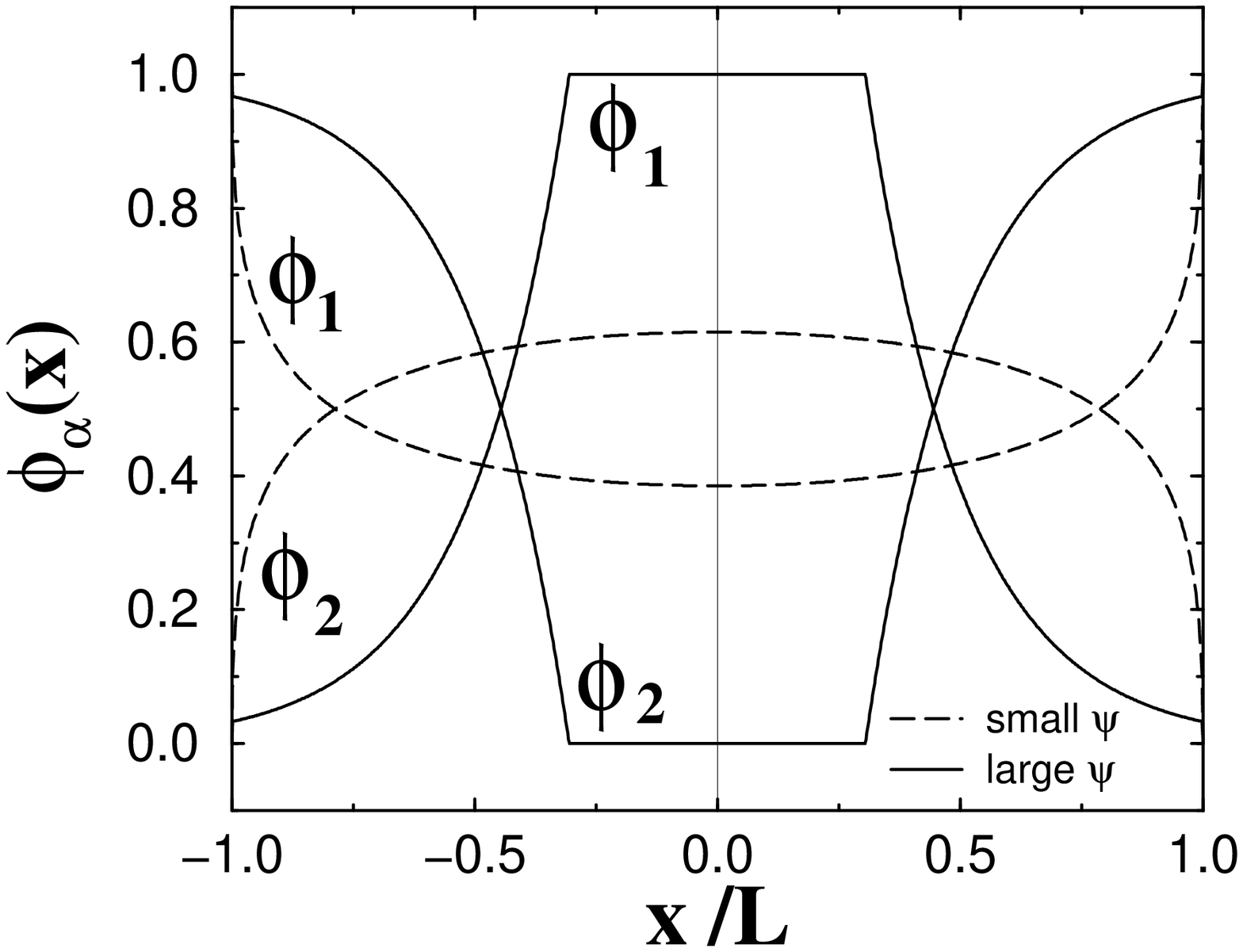}}
}}
\vspace{-.3cm}
\narrowtext
\caption{}
\label{theory}
\end{figure}

Using (\ref{bdg-phi}) and (\ref{linear}) we obtain the profile of the
surface as a function of the rolling species
%
\begin{equation}
\theta(x) - \theta_1(\phi_2) = \frac{- \psi R_2 } {R},
\label{theta}
\end{equation}
where we use that $d_i/\psi x \ll 1$.
The equation for the rolling species of type $1$ is obtained from
Eqs. (\ref{bdg-eq}) and (\ref{theta})
\begin{equation}
\frac{\partial R_1(x)}{\partial x} = \frac{\partial R(x)}{\partial x}
\left ( 1 + \frac{\psi \gamma R_2}{C R} \right ) \frac{R_1}{R^2} .
\end{equation}
By setting $u(x) \equiv R_1(x)/R(x)$ we obtain $u'/[u (1-u)] = - \xi ~
x/[L^2-x^2]$, where $\xi\equiv \gamma \psi/C$, and we arrive to a
power-law form
\begin{equation}
R_1(x) / R(x) = \frac{1}{1+ A (L^2 - x^2)^{-\xi}},
\end{equation}
where the power-law exponent $\xi\sim \psi \approx 0.2$ \cite{makse3}
depends on the degree of difference between the species, and
$A=(\Phi_2/\Phi_1) L^{2\xi}$ is an integration constant obtained from
 $(1/2L)\int_{-L}^L \phi_i(x) dx = \Phi_i$,
where $\Phi_i$ is the fraction of species $i$ at the surface (usually
$\Phi_i=1/2$).  The surface concentration profile
shows a slow power-law decay
\begin{equation}
\phi_1(x) = u [ 1 + \xi (1-u) ].
\label{phi-power}
\end{equation}
Figure \ref{theory} shows the solutions obtained for this case (small
$\psi$). The species $1$--- the smallest grains or the roughest
grains--- are found preferentially at the center of the drum
[$\phi_1(x=0)>\phi_2(x=0)$]. The degree of segregation is small; the
surface concentrations behave very smoothly as a function of the
position in the drum.

According to (\ref{phi-power}), the exponent $\xi$, controls the
intensity of segregation. Since $\xi\sim \psi \propto d_2/d_1$, then
as we increase the size ratio, $\psi$ increases,  and
the linear approximation of the collision functions (\ref{linear})
breaks down.  In this case, strong segregation effects act in the
system.  Segregation happens in the flowing layer as the large grains
are observed to rise to the top of the rolling phase, while the small
grains sink through the gaps left by the large grains; an effect known
as kinematic sieving, free-surface segregation or percolation
\cite{perco,makse3}.  Then the small grains are captured first near
the center of the drum, since they interact with the surface before
than the large grains.
Thus, for large $\psi$ (in terms of size ratios $d_2/d_1>1.5$
\cite{hans}) the capture function of the large grains $b_2$ vanish
near the angle of repose of the small type 1 species: the small
rolling grains screens the interaction of the large rolling grains
with the surface.  We simulate this effect by considering the
following non-linear collision functions \cite{makse2} (valid for
large $\psi$):
\begin{equation}
\begin{array}{llcl}
a_i(\theta)&\equiv& \gamma & \Pi[\theta(x)- \theta_i(\phi_j)] \\
b_i(\theta)&\equiv& \gamma & \Pi[\theta_i(\phi_j)-\theta(x)],
\end{array}
\label{canonical2}
\end{equation}
where $\Pi[z] = 0$, if $z < 0$, and $\Pi[z] = z$, if $z \ge 0$ [Fig.
\ref{drum} (c)].

Next we find the solutions for $x<0$ using (\ref{canonical2}) \cite{same}. 
We consider that
there are $R_1^0$ and $R_2^0$ rolling species of type $1$ and $2$
respectively at the center of the drum ($R_1^0+R_2^0= \sqrt{w/\alpha}
L$, and $R_2^0 / (R_1^0+R_2^0) \simeq \Phi_2$).  Since the collision
functions are defined according to the value of $\theta(x)$ in
comparison to $\theta_i(\phi_2)$, we divide the calculations in two
regions: inner region, where
$\theta_2(\phi_2)<\theta<\theta_1(\phi_2)$, and outer region, where
$\theta<\theta_2(\phi_2)<\theta_1(\phi_2)$.

{\it Inner Region. } If $\theta_2<\theta<\theta_1$, we obtain from
(\ref{bdg-phi})
and (\ref{canonical2}): $\phi_1(x)=1, \phi_2(x)=0$.
Then we obtain the profile of the rolling species using (\ref{bdg-eq})
\begin{equation}
\label{r1}
R_1(x) = \sqrt{\frac{w}{\alpha}} (L^2 - x^2)^{1/2} - R_2^0, \>\>\>\>
R_2(x) = R^0_2.
\end{equation}
The profile of the surface is:
$\theta(x)-\theta_{11} =w x R_1(x) /\gamma.$
This solution is valid when
$\theta(x)>\theta_2(\phi_1=1)=\theta_{21}$, or for $0>x>x_m =[\Phi_2
\sqrt{w\alpha}/\gamma\psi -\sqrt{1-\Phi_2^2+w\alpha/(\gamma\psi)^2}] L
/ [1+w\alpha / (\gamma\psi)^2]$.

The profile of the small rolling grains decays with the distance from
the center as in the case of the total rolling species
Eq. (\ref{total}), since the small grains are the only one effectively
interacting with the surface due to percolation, so that it is a
single species problem.  Next, we find the solution of the problem at
the lower end of the drum.

{\it Outer Region}.  If $\theta<\theta_2<\theta_1$ $(-L \le x \le
x_m)$,
using (\ref{bdg-eq}) and (\ref{canonical2})
%
we obtain the equations for the rolling species
\begin{equation}
\frac{\partial R_1(x)}{\partial x} =- \left ( w x - \gamma \psi R_2
\right ) ~\frac{R_1}{\alpha R(x)^2}.
\label{q1}
\end{equation}
To solve this equation we set $u\equiv R_1(x)/R(x)$, and obtain $
u'/[(1-u) u] = \gamma \psi / [\alpha R(x) ]$ and the solution is
\begin{equation}
{\displaystyle R_1(x) / R(x) = \frac{1}{1+ A \exp[-(x - x_m)/x_s]}},
\end{equation}
where the characteristic length of segregation is $x_s = \sqrt{w
\alpha} L / (\gamma \psi) \sim d_i/\psi$, $A=\Phi_2/\Phi_1$ is an
integration constant obtained by considering the continuity at
$x=x_m$, and we have replaced $\exp\{\sin^{-1}[(x-x_m)/L]\} \simeq
\exp[(x-x_m)/L]$.
The concentration profiles is
\begin{equation}
\phi_1(x) = \exp[(x - x_m)/x_s].
\label{concen2}
\end{equation}
Figure \ref{theory} shows the theoretical profiles found for this case
(large $\psi$).  At the center of the drum we have total segregation
of the small grains [$\phi_1(x=0) = 1$, $\phi_2(x=0) = 0 $].  Then the
concentration of small grains decays exponentially--- with a region of
mixing of characteristic length $x_s$. We also notice that in this
case the profiles depend on $w$ [$x_s\sim \sqrt{w}$] while in the
case of small $\psi$, the profiles (\ref{phi-power}) are independent
of the rotation speed.  Typical values of the phenomenological
constants of the theory have been obtained experimentally in
\cite{makse3}: $\gamma\approx 23$/s, $\alpha\approx 25$/s [see also
Fig. \ref{exp}(b)], and $\psi\approx 0.2$. Then we obtain $x_s\approx
1.6$ cm. These results are in quantitative agreement with experiments
of Cl\'ement {\it et al.} \cite{duran}, who found an exponential decay
in the concentration of particles with a characteristic length of
$\approx 1.7$ cm for size ratio $d_2/d_1$ = 1.5. We also notice that
$x_s$ decreases as $d_2/d_1$ increases, since $\psi\propto d_2/d_1$; a
relationship found in the experiments of \cite{duran}, too.

In sum, we show that the predictions based on the common assumption of
a constant velocity profile of flowing grains are in error. The
profiles predicted by the theory agree with experiments only when a
height-averaged velocity profile is taken into account.  If the 
difference between the species is wide, we find a sharp exponential
segregation in agreement with experiments.  It
would be interesting to test experimentally the transition to the
length scale-free segregation regime when $d_2/d_1$ is decreased below
$\sim 1.5$.  We believe that the present analysis has the potential to
unify the mechanisms underlying segregation in different geometries
such as silos and rotating drums.
Our results are also valid for  granular mixtures poured in
2-D silos:
the different segregation regimes have been observed in 
\cite{bdg,makse2}, and experimentally by
Grasselli and Herrmann \cite{hans}.

I wish to thank T. Boutreux for many discussions, and Coll\`ege de France
for hospitality in initial stages of this work.

\vspace{-.2cm}

FIG. \ref{drum}(a) The thin rotating drum. (b) Linear approximation
(solid lines) to the collision functions (dashed lines) when the
difference between the species is small. (c) Non-linear approximation
(solid lines) to the collision functions when the species differ
appreciably. The collision functions are expected to be continuous
(dashed lines), but we approximate them as discussed in the text.

FIG. \ref{exp}(a) Theoretical prediction (\protect\ref{lambda}) for
$\lambda(w)$ and comparison with experiments of
\protect\cite{nakagawa,khakhar}.
(b) $R(x)$ profiles from theory and experiment
\protect\cite{khakhar}.
We also plot the prediction obtained with the assumption  $v=$const.

FIG. \ref{theory} Theoretical profiles $\phi_i(x)$
for small (dashed lines) and wide (solid lines)
 differences between the species (small and large $\psi$, respectively).

\vspace{-.5cm}

\vspace{-.5cm}

\end{multicols} 

\begin{references}
\bibitem{deviations}
Deviations for $f>20$ rpm may indicate the limit
of validity of the continuum avalanche regime.
\bibitem{parabolic}
The profile based on $v=$const is obtained using 
$v=19.4$ cm/s at $x=0$ \protect\cite{khakhar}.  Even by
rescaling the value of $v$ so that experiment and theory coincide at
$R(x=0)$, we find that the shape of $R(x)$ obtained when $v=$const
does not fit the experiments.
\bibitem{same}
Profiles for $x>0$ are the same 
as  $x<0$ since the grains perform
a solid rotation without deforming.
\end{references}
\end{document}